\documentclass[twocolumn,amsmath,amssymb,prc]{revtex4-1}
\usepackage{graphicx}
\usepackage{color}
\usepackage{bm}

\begin{document}
\title{Variational approach with the superposition
  of the symmetry-restored quasi-particle vacua for nuclear
  shell-model calculations 
}

\author{
  Noritaka Shimizu$^{1}$\footnote{shimizu@cns.s.u-tokyo.ac.jp},
  Yusuke Tsunoda$^1$,
  Yutaka Utsuno$^{2,1}$,
  and
  Takaharu Otsuka$^{3,4,2}$,
}

\affiliation{
  $^1$Center for Nuclear Study, The University of Tokyo,
  7-3-1 Hongo, Bunkyo-ku, Tokyo 113-0033, Japan\\
  $^2$Advanced Science Research Center,
  Japan Atomic Energy Agency, Tokai, Ibaraki 319-1195, Japan \\
  $^3$Department of Physics, The University of Tokyo, 7-3-1 Hongo,
  Bunkyo-ku, Tokyo 113-0033, Japan \\
  $^4$RIKEN Nishina Center, 2-1 Hirosawa, Wako, Saitama 351-0198, Japan \\
}

\begin{abstract}
  We propose a variational calculation scheme utilizing the 
  superposition of the angular-momentum, parity, number projected
  quasiparticle vacua, that is especially suitable 
  for applying to medium-heavy nuclei in shell-model calculations.
  We derive a formula for the energy variance 
  with quasi-particle vacua and apply the energy-variance extrapolation 
  to the present scheme for further precise estimation of
  the exact shell-model energy.
  The validity of the method is presented 
  for the shell-model calculation of
  $^{132}$Ba in the $50 \leq Z,N \leq 82$ model space.
  We also discuss the feasibility of this scheme in the case of 
  the $^{150}$Nd in the $50 \leq Z \leq 82$ and $82 \leq Z \leq 126$ model space
  and demonstrate that its neutrinoless-double-beta-decay matrix element
  is obtained showing good convergence.
\end{abstract}

\maketitle

\section{Introduction}
\label{sec:intro}

Nuclear shell model calculation can describe
any many-body correlations inside the valence shell
on equal footing 
by configuration mixing 
and it is one of the most powerful tools to investigate 
the ground and low-lying excited states of nuclei 
\cite{caurier_rmp,otsuka_rmp}.
However, the number of the configurations to be considered,
namely the dimension
of the shell-model Hamiltonian matrix,
increases explosively depending on the model space
and the number of the active particles, and thus it 
hampers the application of shell-model calculations  to the medium-heavy nuclei
strictly. 
Several shell-model codes have been developed for
massively parallel computations to treat such a large-scale problem 
\cite{mfdn,kshell,bigstick}.
Despite these appreciable efforts, the application
of the conventional shell-model calculation is restricted
by the limitation of available computational resources.
The current feasible $M$-scheme dimension of the  Hamiltonian matrix 
is $O(10^{11})$, which implies that shell-model calculations
are applicable 
only to near semi-magic nuclei in medium-heavy mass region.

In order to overcome this difficulty and to broaden the applicability
of the configuration-mixing framework, 
a lot of efforts have been paid to develop various theoretical frameworks
to obtain shell-model solutions where the conventional Lanczos 
diagonalization method
cannot reach, such as
the projected shell model \cite{proj_sm},
the pair truncation \cite{npa,yoshinaga_npa},
the Monte Carlo shell model (MCSM)
and its extension \cite{mcsm_ppnp,mcsm_ptep},
the VAMPIR approach and its variants \cite{vampir,vampir_ppnp},
the hybrid multi-determinant method \cite{puddu_hmd}, 
the iterative diagonalization algorithm \cite{iter_diag},
the correlated-basis method \cite{corr_sm},
the density matrix renormalization group method \cite{dmrg_sm}, 
the importance truncated shell model \cite{itsm_extrap},
% the double-step truncation scheme \cite{dstep},
and the generator coordinate method (GCM) \cite{hfb_gcm}. 
Recently, the GCM method has been introduced
into the in-medium similarity renormalization
group method to evaluate 
nuclear matrix elements for neutrinoless double-beta decay
\cite{yao_gcm_srg_1,yao_gcm_srg_2,yao_gcm_srg_3}.
Note that the auxiliary-field quantum Monte Carlo approach
can be used to the configuration mixing approach, 
but the realistic shell-model Hamiltonian has
the Fermion sign problem, which restricts
its application to the practical shell-model calculation severely
\cite{smmc,smmc_gammasoft,cp_smmc}.

Among them, the MCSM is one of the most successful
schemes and has been applied
to various mass regions 
\cite{mcsm_ppnp,mcsm_ptep}.
The MCSM wave function is expressed as
a linear combination of the angular-momentum and parity projected
Slater determinants, 
which are determined by the variational and stochastic ways
to minimize the projected energy.  
Introducing the energy-variance extrapolation method to the MCSM
provides us with the more precise estimation
of the exact shell-model energy than the upper limit of
variational method \cite{mcsm_extrap}.
The MCSM has been quite successful in $pf$-shell nuclei \cite{mcsm_ppnp,qmcd_otsuka,ytsunoda_ni}, 
the nuclei around the island of inversion
\cite{utsuno_ioi,ntsunoda_prc,ntsunoda_nature},
an {\it ab initio} approach to light nuclei \cite{liu_mcsm,tabe_mcsm},
and several medium-heavy nuclei \cite{togashi_zr,togashi_sn,otsuka_sm}.
However, in the study of medium-heavy nuclei
where the density of single-particle states per energy
increases and the pairing correlation becomes important \cite{bmpines}, 
a large number of the Slater determinants for the MCSM wave function
are required in principle to describe pair-correlated many-body
wave functions, which often makes 
the precise estimation of exact shell-model physical quantities problematic.
In order to treat such pairing correlation
more efficiently, we introduced
the pair-correlated basis state to the MCSM in Ref.~\cite{ba_mcsm},
although only schematic interactions can be treated in this method.

In the present work, we introduce quasi-particle vacua 
as a replacement of Slater determinants of the MCSM 
to treat various correlations including
pairing correlations efficiently.
We perform the variational calculation
to minimize the energy  after the angular-momentum,
parity, and number projections and superposition.
Hereafter, we call this scheme the quasi-particle vacua shell model (QVSM).
In the same way as the MCSM framework \cite{mcsm_extrap}, we can introduce the
energy-variance extrapolation to overcome the variational limit. 
Note that the importance of the variation
after the number projection  
was discussed in the context of
the Hartree-Fock-Bogoliubov method in Ref.~\cite{egidoring},
and variation after angular-momentum and parity projections 
was achieved in the VAMPIR approach \cite{vampir}.

The evaluation of the nuclear matrix element (NME)
of the neutrinoless double-beta ($0\nu\beta\beta$)
decay is one of the most interesting issues in nuclear structure physics
\cite{avignone08,engel_menendez_bb_review}.
The shell-model calculation is an important model
to estimate the NME precisely since various many-body correlations 
can be included  on an equal footing in the shell-model wave function
\cite{caurier_bb,iwata_bb,dgt_bb,menendez2009,corragio_bb}.
In the present work, we demonstrate that the QVSM 
is useful also to estimate the NME.

This paper is organized as follows:
A form of the variational wave function is introduced
and the feasibility of the variational calculations of the QVSM
is discussed in Sect.~\ref{sec:var}. 
Section \ref{sec:variance} is devoted to the extrapolation
method utilizing the energy variance in the QVSM scheme. 
The applicability of the QVSM to estimate
the $0\nu\beta\beta$-decay NME is discussed in 
Sect.~\ref{sec:nme}.
A summary and future perspectives are given in Sect.~\ref{sec:summary}.
Some derived equations which are required for the present work 
are shown in Appendix.

\section{Variational calculation}
\label{sec:var}

In the QVSM, the variational wave function is defined as
a superposition of the angular-momentum, parity,
and number projected quasi-particle vacua: 
\begin{eqnarray}
  |\Psi_{N_b} \rangle &=&
  \sum_{n=1}^{N_b}\sum_{K=-J}^{J}
  f^{(N_b)}_{nK} P^{J\pi}_{MK} 
  P^Z | \phi_n^{(\pi)} \rangle \otimes P^N | \phi_n^{(\nu)} \rangle 
  \label{eq:qvsmwf}
\end{eqnarray}
where the $P^{J\pi}_{MK}$, $P^Z$, and  $P^N $
are the angular-momentum and parity projector,
the proton number projector, and the neutron number projector, respectively.
$| \phi^{(\pi)}_n \rangle $  ($| \phi^{(\nu)}_n \rangle $) 
denotes the quasi-particle vacuum of protons (neutrons).
$N_b$ is the number of the basis states, or the projected quasi-particle vacua.
$f_{nK}$ is a coefficient
of the linear combination of the basis states
and determined by solving
the generalized eigenvalue problem of
the $(2J+1)N_b\times (2J+1)N_b$
Hamiltonian and norm matrices
in the subspace spanned by the projected
basis states.

The quasi-particle vacuum $| \phi_n \rangle $ is 
parametrized by
complex matrices $U_{ij}^{(n)}$ and $V_{ij}^{(n)}$ as 
\begin{eqnarray}
  \beta^{(n)}_k |\phi_n \rangle &=& 0  \ \ \  \textrm{for  any } k
  \nonumber \\
  \beta^{(n)}_k &=& \sum_i ( V^{(n)*}_{ik} c^\dagger_i + U^{(n)*}_{ik} c_i )
  \label{eq:qv}
\end{eqnarray}
where $\beta_k$ denotes a quasi-particle annihilation operator and 
$c^\dagger_i$ is the creation operator of the single-particle 
orbit $i$ \cite{ringschuck}. 
Note that we do not assume any symmetry for this state.

As $N_b$ increases from 1, the variational parameters $U^{(N_b)}$ and $V^{(N_b)}$  
are determined at every $N_b$ so that the energy expectation value
after the projections and superposition, 
$E_{N_b} = \langle \Psi_{N_b} | H | \Psi_{N_b} \rangle$,
is minimized 
using  the conjugate gradient method \cite{num_recipe} iteratively.
Obeying the variational principle, 
$E_{N_b}$ is the variational upper limit for the exact shell-model energy, 
and $E_{N_b}$ decreases gradually as $N_b$ is increased. 
We stop this iteration when the $E_{N_b}$ converges. 
Unlike the VAMPIR approach \cite{vampir,vampir_ppnp,shapecoex_vampir}, 
we do not include the proton-neutron correlated pair 
for the quasi particle in the present work,
since we aim at investigating neutron-rich nuclei
where the Fermi levels of the protons and neutrons are expected to be apart 
from each other.

To discuss the capability of the QVSM,
we perform the shell-model calculations of $^{132}$Ba
with the SN100PN interaction  \cite{sn100pn}.
The model space is taken as
the $0g_{7/2}$, $1d_{5/2}$, $1d_{3/2}$, $2s_{1/2}$,
and $0h_{11/2}$ orbits  
both for protons and neutrons. 
Its $M$-scheme dimension is $2.0\times 10^{10}$,
and the exact shell-model energy was obtained
by the conventional Lanczos method
with the Oakforest-PACS supercomputer
employing the KSHELL code \cite{kshell}.

\begin{figure}[htbp]
  \includegraphics[scale=0.32]{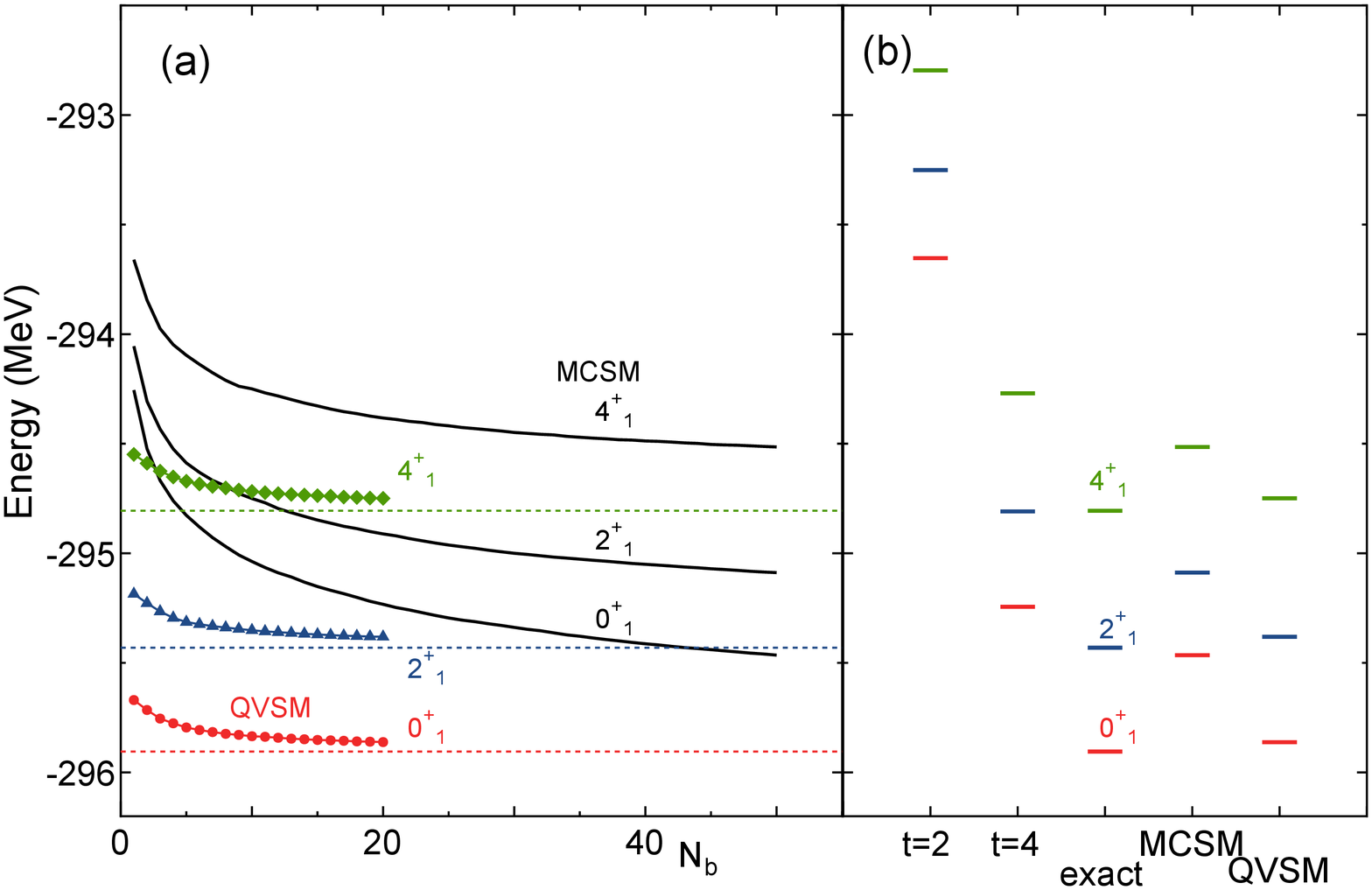}
  \caption{
    (a) Energy expectation values
    $E_{N_b}$ of  the $0^+_1$ (red), $2^+_1$ (blue), $4^+_1$ (green) states
    of $^{132}$Ba by the QVSM and the MCSM
    as a function of the number of the basis states $N_b$.
    The horizontal dotted lines show the exact shell-model energies.
    (b) The $0^+_1$, $2^+_1$, $4^+_1$ energies of $^{132}$Ba
    obtained by the conventional Lanczos method
    with $t=2$ truncation, by that with $t=4$ truncation,
    by the exact calculation without truncation, 
    by the MCSM with 50 basis states, and by the QVSM with 20 basis states
    are shown from left to right.
  }
  \label{fig:ba132j024}
\end{figure}

Figure \ref{fig:ba132j024}(a) shows the QVSM energy 
as a function of the number of the basis states $N_b$,
which is defined in Eq.(\ref{eq:qvsmwf}).
As $N_b$ increases the QVSM energy comes down
and the energy converges rapidly and approaches the exact values
shown in Fig.~\ref{fig:ba132j024}(b).
Even at  $N_b=1$, the QVSM energy is closer to the exact one
than those of $t$-particle $t$-hole truncations and that of the MCSM.

The solid black lines in Fig. \ref{fig:ba132j024}(a) 
also show the MCSM energy expectation values as a function
of the number of the basis states, $N_b$.
The MCSM wave function \cite{mcsm_ptep} is 
defined as a linear combination of the angular-momentum and 
parity projected deformed Slater determinants as 
\begin{eqnarray}
  |\Psi_{N_b} \rangle &=&
  \sum_{n=1}^{N_b}\sum_{K=-J}^{J}
  f^{(N_b)}_{nK} P^{J\pi}_{MK} | D_n^{(\pi)} \rangle
  \otimes | D_n^{(\nu)} \rangle
  \label{eq:mcsmwf}
\end{eqnarray}
where $| D_n^{(\pi)}\rangle$  and $| D_n^{(\nu)}\rangle$
are the deformed Slater determinants for protons and neutrons, respectively.
The number projection is not necessary since
a Slater determinant is an eigenstate of the number operator.
The Slater determinant $|D_n\rangle$ 
is parametrized by the complex matrix $D_n$, 
which is determined to minimize the energy eigenvalue
in the same way as the QVSM.
Since Slater determinants cannot describe pairing correlations
efficiently, the MCSM energy converges rather slowly 
in comparison with the QVSM.

For comparison, Figure \ref{fig:ba132j024} (b)
shows the energies obtained by the QVSM,
the MCSM, and the conventional Lanczos diagonalization method 
in truncated spaces.
The conventional Lanczos method was performed with 
the truncated space restricting up to $t$-particle
$t$-hole excitations across the $Z=N=64$ shell gap
from the filling configuration.
The truncation scheme is taken as $t=2$ 
($2.3 \times 10^8$ $M$-scheme dimension),
$t=4$ ($3.4 \times 10^9$ $M$-scheme dimension), 
and the full space without truncation for the exact energy.
The $t=2$ ($t=4$) energy is 2.3 MeV (0.7 MeV) higher
than the exact one. 
The rightmost part of Fig. \ref{fig:ba132j024} (b)
shows the results of the MCSM with $N_b=50$
and the QVSM with $N_b=20$.
The MCSM result overcomes the truncated results, 
but it is still 440 keV higher than
the exact one and the MCSM underestimates the excitation energies
100 keV.
The QVSM agrees with the exact one quite well 
within 50 keV.
This small gap between the QVSM and exact ones
can be filled by the energy-variance
extrapolation method, 
which will be discussed in the next section.

% The computational cost of the variational calculations
% to determine $N_b$-th basis state
% in the QVSM is roughly estimated
% as a product of the number of mesh points of the number projection
% and the cost of the MCSM to determine $N_b$-th basis state.
While the QVSM converges as a function of $N_b$
apparently better than the MCSM,
the computational cost of the QVSM with the same $N_b$
is heavier than that of the MCSM
because of the necessity of the number projection.
In practice, the total computational of 
the QVSM whose result is shown in Fig. \ref{fig:ba132j024}
is about 10 times heavier than that of the MCSM.
Even considering such difference, the QVSM is 
more efficient than the MCSM in the $^{132}$Ba case 
since the QVSM energy with $N_b=1$ is already lower than 
the MCSM energy with $N_b=50$.

\begin{figure}[htbp]
  \includegraphics[scale=0.4]{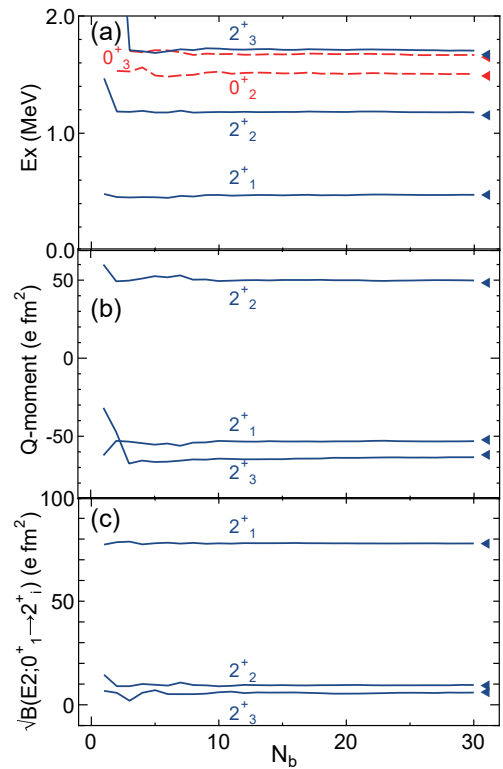}
  \caption{
    (a) Excitation energies of the
    $0^+_2$ and $0^+_3$ states (red dashed lines),
    and $2^+_1$, $2^+_2$, and $2^+_3$ states (blue solid lines)
    of $^{132}$Ba against the number of the basis states $N_b$.
    (b) Quadrupole moments of the $2^+_1$, $2^+_2$, and $2^+_3$ states.
    (c) Square root of the B(E2) transition probabilities from the
    ground state to the $2^+_1$, $2^+_2$, and $2^+_3$ states.
    The triangles at the rightmost side denote
    the exact values. 
  }
  \label{fig:ba132n3-ex}
\end{figure}

For proving the feasibility of the QVSM to treat 
the nonyrast states and other physical quantities, 
we performed the variational calculations to
obtain 
the lowest three $0^+$ and $2^+$ states of
$^{132}$Ba.
We performed the variational calculations
so that the summation of the
lowest three energy expectation values
is minimized.
Figure  \ref{fig:ba132n3-ex} (a)
shows the excitation energies of the $0^+_2$, $0^+_3$,
$2^+_1$, $2^+_2$, and $2^+_3$ states as a function
of the number of the basis states $N_b$.
The extrapolation procedure is not required 
since these excitation energies converge quite rapidly
at $N_b \simeq 10$ and agree with the exact values.
Figures  \ref{fig:ba132n3-ex} (b) and (c)
show the quadrupole moments and the square root
of the B(E2) transition probabilities
with the effective charges  $(e_p,e_n) = (1.5, 0.5)e$.
These observables also show good convergence patterns
and converge at $N_b\simeq 10$.
Note that although $B(E2;0^+_1 \rightarrow 2^+_2) $ and
$B(E2;0^+_1 \rightarrow 2^+_3) $ are quite small in comparison
with the large $B(E2;0^+_1 \rightarrow 2^+_1)$ value,
these three $B(E2)$ values converge rapidly at the same pace.

\section{Energy variance extrapolation}
\label{sec:variance}

The variational calculation discussed in the previous section
gives us only the variational upper limit to the exact shell-model energy.
In order to estimate the exact energy more precisely,
we here introduce the extrapolation method employing the energy variance.
The energy-variance extrapolation was proposed in condensed matter physics
\cite{sorella}, 
and  was introduced to the nuclear shell-model calculations
in Ref.~\cite{mizusaki_imada_extrap}.
Since then it was applied to various schemes
\cite{mcsm_extrap,itsm_extrap,puddu_variance,extrap_ncsm,vmcsm}.

The energy variance of the variational wave functions is defined as
\begin{eqnarray}
  \langle \Delta H^2 \rangle_{N_b}
  &=& \langle \Psi_{N_b}| H^2|\Psi_{N_b} \rangle
  -   \langle \Psi_{N_b}| H  |\Psi_{N_b} \rangle^2 . 
\end{eqnarray}
A formula to compute the energy variance in the quasi-particle vacua
is shown in Appendix \ref{sec:app_variance}.
By utilizing the fact that the energy variance is zero
if $|\Psi_{N_b} \rangle$ is
the exact shell-model eigenstate,
the variance expectation value
is not only an indicator for the approximation,
but also can be used for the estimation of the exact
energy eigenvalue by extrapolation.
As $N_b$ increases,
the QVSM wave function approaches the exact one
and the corresponding variance approaches zero.
In the extrapolation scheme, 
we plot the energy $E_{N_b}$ against the variance
$\langle \Delta H^2 \rangle_{N_b}$,
which is called a variance-energy plot hereafter.
On the plot, as $N_b$ increases the point is expected
to approach the $y$ axis,
namely $\langle \Delta H^2 \rangle_{N_b}=0$, gradually.
The extrapolated energy is the $y$ intercept
of the curve fitted for these points.

\begin{figure}[htbp]
  \includegraphics[scale=0.4]{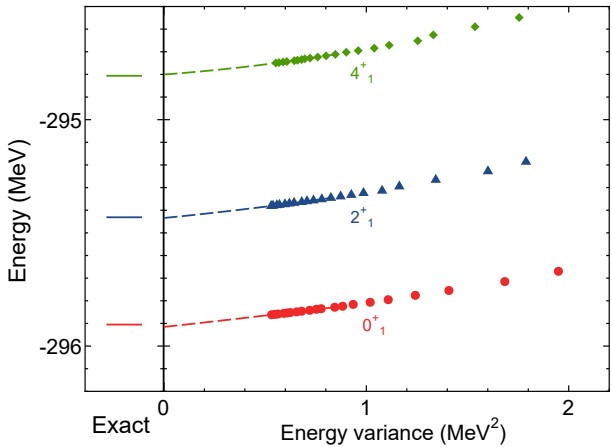}
  \caption{
    Variance-energy plot of $^{132}$Ba. 
    The red circles, blue triangles, and green diamonds denote
    the energy expectation values against the energy variance
    of the $0^+_1$, $2^+_1$, and $4^+_1$ states, respectively,
    obtained by the QVSM.
    The dashed curves are drawn to be fitted
    for the last 12 points of the QVSM results with
    a 2nd-order polynomial.
    and their $y$-intercepts are the extrapolated values.
    The left panel shows the exact shell-model
    energies.
  }
  \label{fig:eve-ba132j024}
\end{figure}

Figure \ref{fig:eve-ba132j024} shows the variance-energy plot
of the QVSM wave functions of $^{132}$Ba,
which are the same as the case in Fig. \ref{fig:ba132j024}.
The last 12 points are used for the 2nd-order polynomial fit.
The extrapolated values, which are the $y$ intercepts of the fitted lines, 
and the exact shell-model energies
agree with each other quite well within a 10-keV difference.

\begin{figure}[htbp]
  \includegraphics[scale=0.4]{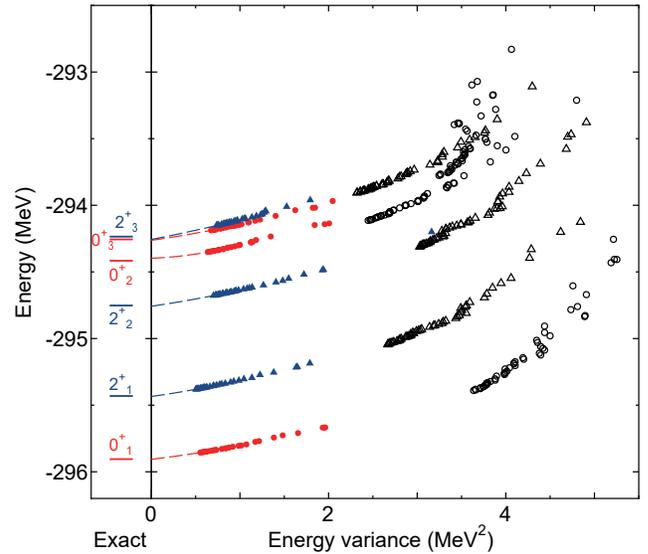}
  \caption{
    Variance-energy plot of $^{132}$Ba. 
    The red circles (blue triangles) denote 
    the lowest three $0^+$ ($2^+$) states obtained
    by the QVSM. The black open circles (triangles)
    denote the energies and energy variances of
    the $0^+$ ($2^+$) states of the MCSM. 
    The dashed curves are drawn to be fitted
    for the last 20 points of the QVSM results with
    a 2nd-order polynomial.
    and their $y$-intercepts are the extrapolated values.
    The left panel shows the exact shell-model
    energies.
  }
  \label{fig:eve-ba132n3}
\end{figure}

Figure \ref{fig:eve-ba132n3} shows
variance-energy plots of the QVSM and the MCSM, 
whose wave functions are the same as the
case of Fig.~\ref{fig:ba132n3-ex}.
While the left panel shows the exact shell-model
energies
the $y$-intercepts of the fitted curves are the
extrapolated values of the QVSM results.
The extrapolated energies and the exact shell-model
energies obtained by the conventional Lanczos method
agree quite well within a 20-keV difference.

%jjlast 20 points?

\begin{figure}[htbp]
  \includegraphics[scale=0.4]{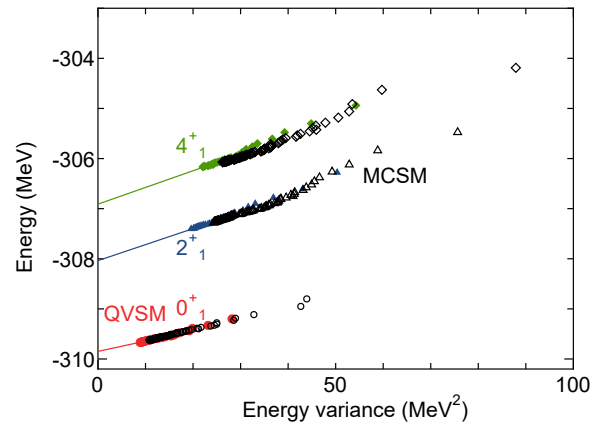}
  \caption{
    Variance-energy plot of $^{68}$Ni with
    the A3DA interaction \cite{ytsunoda_ni}.
    The red filled circles, blue filled triangles,
    and green filled diamonds are the QVSM results of
    the $0^+_1$, $2^+_1$, and $4^+_1$ states, respectively.
    The QVSM is obtained with $N_b=30$.
    The black symbols are the corresponding MCSM results 
    with 120 basis states. 
    The solid lines denote the fitted curves by a 1st-order
    polynomial for the last 13 QVSM points.
  }
  \label{fig:ni68-ev-n1}
\end{figure}

One of the major achievements of the MCSM is
to reveal the exotic structure
of neutron-rich nuclei around $^{68}$Ni \cite{ytsunoda_ni}.
For further comparison of the QVSM and the MCSM,
we show the variance-energy plot of the $^{68}$Ni with
the A3DA interaction \cite{ytsunoda_ni} and the model space
consisting of the $pf$ shell, $0g_{9/2}$, and $1d_{5/2}$
orbits in Fig.~\ref{fig:ni68-ev-n1}.
The $M$-scheme dimension of this system is
$5.2 \times 10^{15}$, which is beyond the current feasibility
of the conventional Lanczos method even now.
The points of the MCSM and the points of the QVSM
show a similar tendency.
Although the QVSM with the 30 basis states provides
us with the lower variational energy 
than that of the MCSM with the 120 basis states, 
the MCSM is advantageous in terms of the computation time
since the number projection is not needed for the MCSM. 
According to our numerical experiments, 
the MCSM is more efficient 
for lighter nuclei such as $pf$-shell nuclei, 
while the QVSM is expected to be more efficient and
to converge faster  
in medium-heavy mass region beyond the $N=Z=50$ gap such as $^{132}$Ba.
The feasibility and application of the QVSM to $^{150}$Nd
and the evaluation of its $0\nu\beta\beta$-decay NME
will be discussed in the next section.

\section{Neutrinoless double beta decay matrix elements}
\label{sec:nme}

Here we focus on the convergence property 
of a NME of $0\nu\beta\beta$ decay in the QVSM and the MCSM.
The $0\nu\beta\beta$-decay NME with the closure approximation
is obtained as 
\begin{equation}
  M^{0\nu} = \langle 0^+_f |\hat{O}|  0^+_i \rangle
  = M^{0\nu}_{GT} - \frac{g_V^2}{g_A^2} M^{0\nu}_{F}
  + M^{0\nu}_{T}
\end{equation}
where $\hat{O}$ is the operator to annihilate two neutrons
and create two protons with neutrino potential.
$M^{0\nu}_{GT}$, $M^{0\nu}_{F}$ $M^{0\nu}_{T}$
denote the Gamow-Teller type, Fermi type, tensor type
terms classified according to spin structure of the operator,
respectively \cite{engel_menendez_bb_review}.
$|0^+_i\rangle$ and $|0^+_f\rangle$ are the ground states
of the parent and daughter nuclei.
$g_V$ and $g_A$ are vector and axial-vector coupling constants
and are taken as 1.0 and 1.27, respectively.
In the present work, a factor of short range correlation
for the NME is omitted for simplicity and 
the average energy of the closure approximation is
taken from the empirical formula 
$E_\textrm{av}=1.12\sqrt{A}$ MeV \cite{haxton}.

As a benchmark test, we perform the shell-model calculation
of $^{76}$Ge and $^{76}$Se with JUN45 interaction \cite{jun45}
with the model space consisting of the $0f_{5/2}$,
$1p_{3/2}$, $1p_{1/2}$, and $0g_{9/2}$ orbits both for protons and neutrons
and evaluate the $0\nu\beta\beta$-decay NME.
% Since its $M$-scheme dimension is $1.7\times10^7$,
Since the $M$-scheme dimension of $^{76}$Se is $6.8\times10^8$,
the exact shell-model value is obtained by the conventional Lanczos method
more easily than the case of $^{132}$Ba.

\begin{figure}[htbp]
  \includegraphics[scale=0.4]{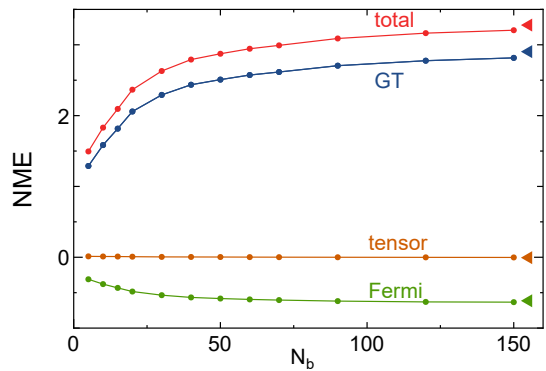}
  \caption{
    $0\nu\beta\beta$-decay NME of $^{76}$Ge obtained by the MCSM.
    These values are shown as a function 
    against the number of the basis states $N_b$.
    The red, blue, green, orange lines with the solid circles
    denote the total, GT-type, Fermi-type, and tensor-type
    NMEs, respectively.
    The circles show the MCSM values with
    $N_b=5,10,15,20,30,40,50,60,70,90,120,$ and 150.
    The exact shell-model values are shown as the triangles at the rightmost.
  }
  \label{fig:nme-ge76-mcsm}
\end{figure}

Figure \ref{fig:nme-ge76-mcsm} shows the NME
obtained by the MCSM against the number of the basis states $N_b$.
The MCSM calculation is performed up to $N_b=150$.
% The convergence of the NME value is so slow 
% that the extrapolation technique is difficult to be applied. 
The NME of the MCSM shows
quite slow convergence as a function of the number of the
basis states 
and the extrapolation using the MCSM results to the exact solution
seems to be difficult. 
Since a $0\nu\beta\beta$-decay NME is 
sensitive to pairing correlations 
\cite{caurier2008,menendez2009,dgt_bb},
the QVSM scheme is expected to be advantageous
over the MCSM.
To evaluate the NME using the linear combination of 
quasi-particle vacua as a wave function
was also discussed in the context of the
generator coordinate method
\cite{yao_gcm_srg_1,yao_gcm_srg_2,yao_gcm_srg_3,gcm_bb,rodriguez_150Nd}.

\begin{figure}[htbp]
  \includegraphics[scale=0.4]{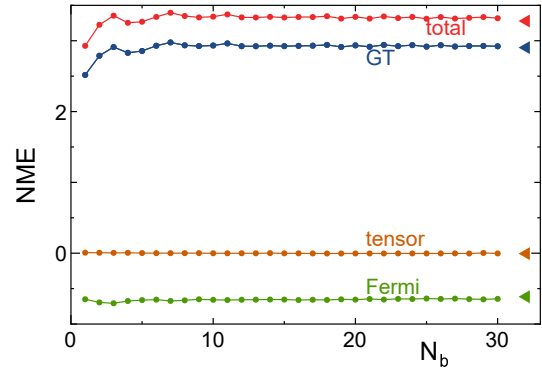}
  \caption{
    $0\nu\beta\beta$-decay NME of the $^{76}$Ge 
    against $N_b$ obtained by the QVSM.
    The exact values are shown as the triangles at the rightmost.
  }
  \label{fig:nme-ge76-qv}
\end{figure}

Figure \ref{fig:nme-ge76-qv} shows the NMEs obtained by the QVSM.
The NMEs of the QVSM converge quite fast and agree well with the
exact shell-model values.
While the total NME of the MCSM is too small at $N_b=1$,
the NMEs of the QVSM are close to the exact one even at  $N_b=1$.
It implies that the efficient treatment of the pairing correlation
is essential for the estimation of the $0\nu\beta\beta$ NMEs.

We evaluate the NMEs of the $0\nu\beta\beta$ decay of $^{150}$Nd
by the QVSM and the MCSM.
We adopt the Kuo-Herling interaction \cite{khhe}
with the model space consisting of
$0g_{7/2}$, $1d_{5/2}$, $1d_{3/2}$, $2s_{1/2}$, and $0h_{11/2}$
orbits for protons and
$0h_{9/2}$, $1f_{7/2}$, $1f_{5/2}$, $2p_{3/2}$, $2p_{1/2}$, and $0i_{13/2}$
orbits for neutrons with the $^{132}$Sn inert core. 
The $M$-scheme dimension of $^{150}$Nd 
is $2.2\times 10^{14}$, far beyond the current limitation
of the conventional Lanczos method.
$^{150}$Nd is one of nuclei whose NME has not been
evaluated by reliable shell-model calculations among major
double-beta-decay nuclei used for $0\nu\beta\beta$-decay 
search experiments \cite{engel_menendez_bb_review}. 

\begin{figure}[htbp]
  \includegraphics[scale=0.45]{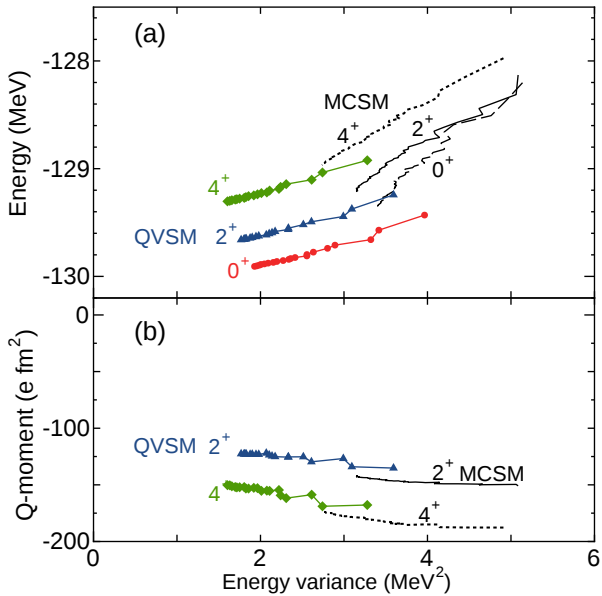}
  \caption{
    Energy expectation values (a) and
    spectroscopic quadrupole moments (b) of the
    yrast states of $^{150}$Nd obtained
    by the QVSM with $N_b=20$ and the MCSM with $N_b=100$
    against the energy variance.
    The red circles, blue triangles, and green
    diamonds denote the QVSM results of 
    the $0^+$,  $2^+$ and $4^+$ states, respectively.
    The dashed, solid, and dotted lines
    are the MCSM results of 
    the $0^+$,  $2^+$ and $4^+$ states, respectively.
    The effective charges are taken as $(e_p,e_n) = (1.58,0.85)e$.
  }
  \label{fig:v-q-Nd}
\end{figure}

\begin{figure}[htbp]
  \includegraphics[scale=0.45]{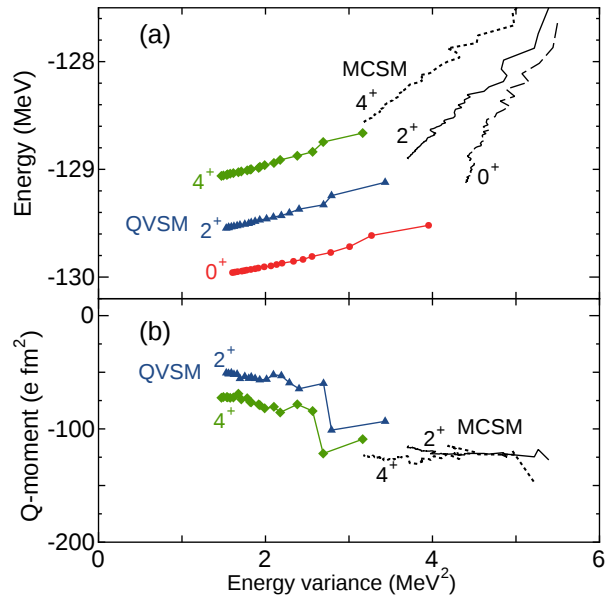}
  \caption{
    Energy expectation values  (a) and
    spectroscopic quadrupole moments (b) of the
    yrast states of $^{150}$Sm obtained
    by the QVSM and the MCSM against the energy variance.
    See caption of Fig.~\ref{fig:v-q-Nd} for details.
  }
  \label{fig:v-q-Sm}
\end{figure}

Figures \ref{fig:v-q-Nd} (a) and \ref{fig:v-q-Sm} (a)
show the variance-energy plots of the  $0^+$,  $2^+$ and $4^+$
states of $^{150}$Nd  and  $^{150}$Sm, respectively.
These are obtained by the QVSM with $N_b=20$
and the MCSM with $N_b=100$.
The QVSM energies are  about 1 MeV lower, and therefore better,
than those of the MCSM, 
While the MCSM results do not reach the convergence even at $N_b=100$,
the QVSM results show good convergence. 
The $0^+$, $2^+$, and $4^+$ energies of the QVSM converge 
in a similar way, which means the excitation energies
converge stably.
Figures \ref{fig:v-q-Nd} (b) and \ref{fig:v-q-Sm} (b)
show the spectroscopic quadrupole moments of the QVSM
and the MCSM results against the energy variance.
The QVSM results show stable convergence, 
while the MCSM overestimates the deformation.

\begin{figure}[htbp]
  \includegraphics[scale=0.35]{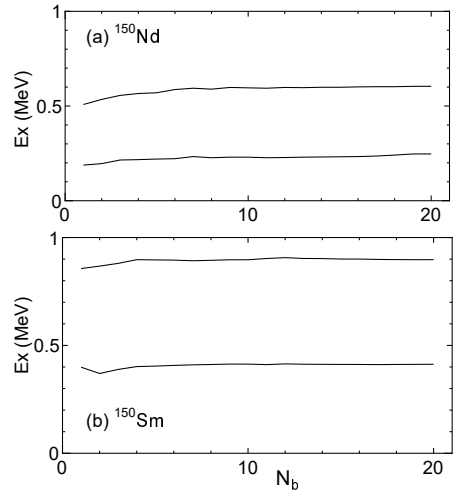}
  \caption{
    Excitation energies of the
    $2^+$ and $4^+$ states of (a) $^{150}$Nd
    and (b) $^{150}$Sm as a function of
    the QVSM basis states.
  }
  \label{fig:ex-Nd-Sm}
\end{figure}

% Figure \ref{fig:ex-q-Nd}  shows the
% excitation energies of $^{150}$Nd 
% obtained by the QVSM and the MCSM. 
% The shell-model energies of the QVSM are
% about 1 MeV lower, and therefore better, than those of the MCSM, 
% and converge well as a function
% of $N_b$. 
Figure \ref{fig:ex-Nd-Sm} shows the
excitation energies of $^{150}$Nd and 
its daughter nucleus, $^{150}$Sm, obtained by the QVSM. 
The energies of the QVSM converge well as a function
of $N_b$.
The QVSM result of the $2^+$ energy of the $^{150}$Nd is
240 keV which is larger than the experimental value, 130 keV.
Besides, the quadrupole moment by the QVSM is
$-1.2$ eb, 
which shows smaller deformation
than the experimental value, $-2.0(5)$ eb.
It may indicate that the larger model space is required
to describe the large quadrupole deformation of $^{150}$Nd,
which is also indicated by Refs.~\cite{Yao_PRC91,otsuka_sm}.
% it is sufficient for the present benchmark test, nevertheless.
% The excitation energies and the quadrupole moments obtained
% by the MCSM increase quite gradually as the number of the
% basis states increases.

% \begin{figure}[htbp]
% %  \includegraphics[scale=0.4]{Nd150-ev-b.eps}
%   \includegraphics[scale=0.4]{Sm150-EQ.eps}
%   \caption{
%     QVSM and MCSM results of the
%     $0^+$, $2^+$ and $4^+$ states of $^{150}$Sm
%     as a function of the number of 
%     the basis states.
%     See caption of Fig.~\ref{fig:ex-q-Nd} for details.
%   }
%   \label{fig:ex-q-Sm}
% \end{figure}

% % 150Sm daugher (Z=62, N=88)
% Figure \ref{fig:ex-q-Sm} shows 
% the theoretical results of the $0\nu\beta\beta$-decay
% daughter nucleus, $^{150}$Sm.
% The QVSM shows apparently better than
% the MCSM in terms of the variational principle. 
% The excitation energies and quadrupole moment obtained by the QVSM
% show good convergence and large deviations from those
% by the MCSM.

\begin{figure}[htbp]
  \includegraphics[scale=0.4]{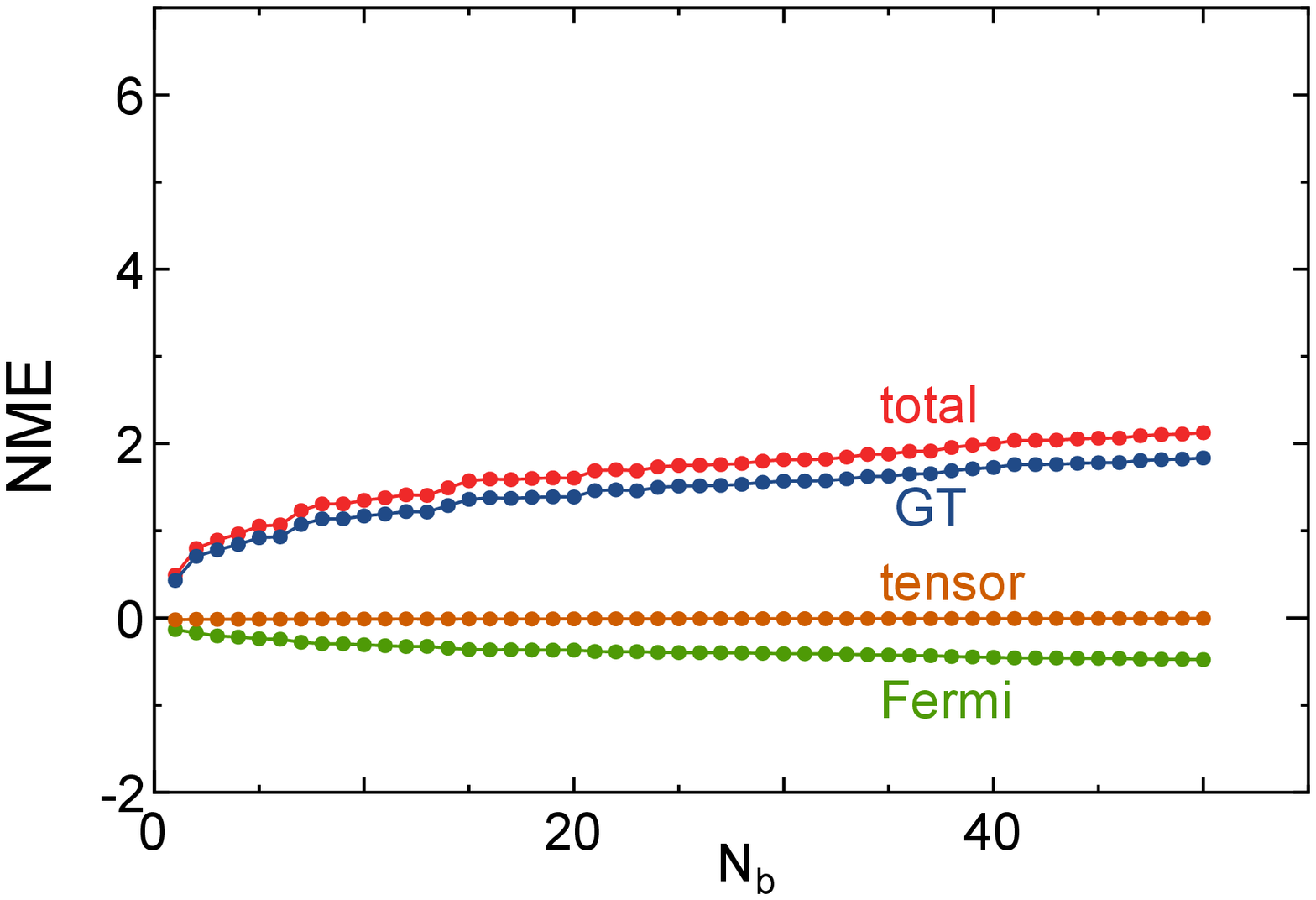}
  \caption{
    $0\nu\beta\beta$-decay NME of the $^{150}$Nd
    obtained by the MCSM 
    against the number of the basis states $N_b$.
    The red, blue, green, orange lines with the solid circles
    denote the total, GT-type, Fermi-type, and tensor-type
    NMEs, respectively.
  }
  \label{fig:nme-Nd150-mcsm}
\end{figure}

\begin{figure}[htbp]
  \includegraphics[scale=0.4]{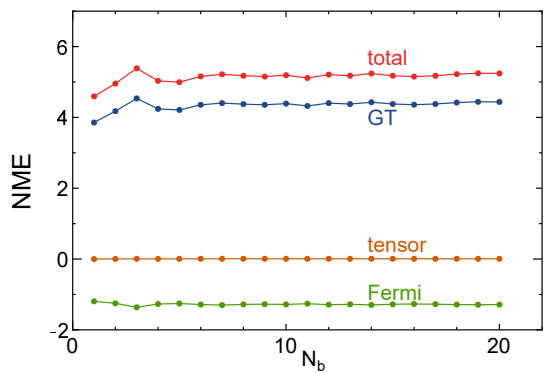}
  \caption{
    $0\nu\beta\beta$-decay NME of the $^{150}$Nd
    obtained by the QVSM
    against the number of the basis states $N_b$.
    The red, blue, green, orange lines with the solid circles
    denote the total, GT-type, Fermi-type, and tensor-type
    NMEs, respectively.
  }
  \label{fig:nme-Nd-Sm}
\end{figure}

Figure \ref{fig:nme-Nd150-mcsm} shows the NME of the MCSM
against the number of the basis states.
The convergence of the NME is quite slow and it is difficult
to estimate the converged value. 
Figure  \ref{fig:nme-Nd-Sm} shows 
the NME values by the QVSM. 
The NMEs converge quite rapidly in contrast
to the MCSM case in Fig. \ref{fig:nme-Nd150-mcsm},
and these values do not change where $N_b$ is beyond 10
owing to the efficient description by the quasi-particle-vacuum basis states.
Although this NME value is not conclusive since
the present study overestimates the excitation energies, 
it is worth comparing it with previous works briefly.
The total NME of the current work is 5.2, which is about twice larger than
other proceeding results of the quasi-particle random phase approximation
and several other approaches \cite{QRPA_CH}.
The NME values given by the latest 
generator-coordinate method based on the relativistic
energy density functional are 5.6 \cite{Yao_PRC91}
and 5.2 \cite{YaoEngel_PRC94},
which are close to the present result.
% It was suggested that the large quadrupole deformation
% contributes to suppress the NME \cite{rodriguez_150Nd,FangFaessler2010}.
It was suggested that the large 
difference between the deformations of the initial and final states 
would suppress the NME
\cite{song_RMF_150Nd,rodriguez_150Nd,FangFaessler2010}.
Further investigation by extending the model space
is ongoing to evaluate the effect of
the large quadrupole deformation and its differences between
the initial and final states appropriately.

\section{Summary}
\label{sec:summary}

We have developed a variational method after
the superposition of the fully projected quasi-particle vacua,
named the quasi-particle vacua shell model (QVSM),
which is an extension of the MCSM.
We apply the energy-variance extrapolation method to the QVSM
and demonstrated that it works quite well 
to estimate the shell-model energies of $^{132}$Ba with the SN100PN interaction.
The excitation energies and other observables
such as the quadrupole moment and $B(E2)$ transition
probabilities by the QVSM converge 
quite rapidly as a function of the number of
the basis states.
Since the QVSM wave function is
expected to include many-body correlations such as pairing correlations
efficiently, 
it works well in nuclei heavier than Sn isotopes, 
while the MCSM is efficient enough in lighter-mass region such as $^{68}$Ni
in terms of computational resources. 

We have demonstrated that the NME values of the QVSM against $N_b$ show
fast convergence. 
The feasibility of the QVSM to evaluate the $0\nu\beta\beta$-decay NME
of $^{150}$Nd is validated.
Since the shell-model result of the Kuo-Herling interaction overestimates
the experimental $2^+$ and $4^+$ excitation energies of
$^{150}$Nd and $^{150}$Sm,
further investigation is anticipated to conclude
the NME values by shell-model calculations.
We also plan to evaluate the NMEs of double-beta-decay nuclei
in medium-heavy mass region 
such as $^{136}$Xe \cite{kamland} and $^{100}$Mo, which will be used
for the next-generation $0\nu\beta\beta$-decay
search experiment \cite{cupid}.

This proof-of-the-principle study
opens a way to investigate the medium-heavy nuclei
with configuration mixing 
utilizing nuclear shell-model calculations.
The application of the present scheme 
to odd nuclei would be rather straightforward
and is under progress.

\section*{Acknowledgment}
\label{sec:ack}

NS acknowledges Drs. Takashi Abe, Javier Men\'{e}ndez,
Takahiro Mizusaki, and Kota Yanase
for fruitful discussions.
This research used computational resources of the supercomputer Fugaku
(The evaluation environment in the trial phase) provided
by the RIKEN Center for Computational Science
through the HPCI System Research project (Project ID:hp200130), 
Oakforest-PACS supercomputer (hp200130, hp190160, xg18i035),
and CX400 supercomputer of Nagoya University (hp160146).
The authors acknowledge valuable supports 
by ``Priority Issue on post-K computer''
(Elucidation of the Fundamental Laws and Evolution of the Universe)
and ``Program for Promoting Researches on the Supercomputer Fugaku''
(Simulation for basic science: from fundamental laws of particles
to creation of nuclei), MEXT, Japan.

\appendix

\section{Matrix elements between different quasi-particle vacua}

In this appendix, we briefly show some
equations which are required for the QVSM scheme.
Since the QVSM wave function defined in Eq.(\ref{eq:qvsmwf})
is written
as a linear combination of the projected quasi-particle vacua, 
we need the equations to compute the overlap, 
the Hamiltonian matrix elements,
and the energy gradient between two different
quasi-particle vacua.
In addition, we firstly derive the equation of 
the matrix elements of the Hamiltonian squared 
for the energy-variance extrapolation technique.

The overlap between two different quasi-particle vacua
is computed by the Neergard-Wust method \cite{neergardwust}
in the present work for efficient computations, while
more elegant formula employing the Pfaffian was suggested
\cite{robledo_pfaffian}.
Some overlap formulae for odd-mass case
were proposed \cite{AvezBender,bertsch_robledo,oimizusaki_odd}.

The shell-model Hamiltonian is defined as
\begin{equation}
  H = \sum_{ij}t_{ij}c^\dagger_i c_j
  + \frac14 \sum_{ijkl} v_{ijkl}c^\dagger_i c^\dagger_j c_l c_k
\end{equation}
where $t$ and $v$ are the coefficients of
the one-body and two-body interactions, respectively.
$v$ is Hermitian ($v_{ijkl}=v_{klij}$)
and anti-symmetrized ($v_{ijkl}=-v_{jikl}=-v_{ijlk}=v_{jilk}$).

\subsection{Hamiltonian matrix elements}
\label{sec:mele}.

We show equations for the 
energy expectation values of the QVSM wave function. 
The QVSM wave function is a linear combination of the
angular-momentum, parity, number projected quasi-particle vacua. 
In numerical calculations, the angular-momentum projector
is calculated 
as the summation of the discretized Euler angles
$\Omega=(\alpha,\beta,\gamma)$, 
and the parity projection is performed utilizing
the parity-conversion operator $\Pi$ \cite{mcsm_ptep}:
\begin{eqnarray}
  P^{J\pi}_{MK}
  &=&  \frac{1 + \pi \Pi}{2}\frac{2J+1}{8\pi^2}
  \int d\Omega \  D^{J*}_{MK}(\Omega)
      e^{i\alpha J_z} e^{i\beta J_y} e^{i\gamma J_z}
      \nonumber \\
  &\simeq& \sum_a W_a^{JMK\pi} R_a
\end{eqnarray}
with 
\begin{eqnarray}
  R_a
  &=& e^{i\alpha_a J_z} e^{i\beta_a J_y} e^{i\gamma_a J_z}
      \Pi^{(\delta_a)} ,
      \\
  W_a^{JMK\pi}
  &=& \frac{2J+1}{8\pi^2}
      D^{J*}_{MK}(\alpha_a,\beta_a,\gamma_a)   \pi^{(\delta_a)} w_a ,
      \nonumber 
\end{eqnarray}
$\pi^{(1)}=\frac12, \pi^{(2)}=\frac{\pi}{2}, \Pi^{(1)}=1, \Pi^{(2)}=\Pi$,
and $a=(\alpha_a,\beta_a,\gamma_a,\delta_a)$ \cite{mcsm_ptep}.
$a$ and $w_a$ are a set of discretized mesh points
and its corresponding weight for the summation, 
and are determined by the Gaussian quadrature \cite{num_recipe}. 
The proton number projector is also computed as
\begin{equation}
  P^{Z} = \frac{1}{2\pi}  \int_0^{2\pi} e^{i\phi (N^{(\pi)}-Z)} d\phi
  \simeq \sum_{b=1}^L W_b^{(Z)} R^{(Z)}_b
\end{equation}
where $N^{(\pi)}$ denotes the proton number operator
and 
$R^{(Z)}_b= e^{2\pi i N^{(\pi)}b/L} $,
$W_b^{(Z)} = \frac1L  e^{-2\pi i Zb/L}$.
The neutron number projector is defined
in the same way as the proton case.

Thus, the coefficient of the QVSM wave function $f^{(N_b)}_{iK}$
in Eq.(\ref{eq:qvsmwf})
and its energy expectation value are
obtained by solving the generalized eigenvalue problem,
or the Hill-Wheeler-Griffin equation \cite{hillwheeler}, 
\begin{equation}
  \sum_{j=1}^{N_b}
  \sum_{K'=-J}^J \left( H_{iK,jK'}- E_{N_b} N_{iK,jK'} \right) f^{(N_b)}_{jK'} = 0
    \label{eq:geneig}
\end{equation}
with 
\begin{eqnarray}
  H_{iK,jK'}
  &=& \sum_{a} W_a^{JKK'\pi}
      \langle \phi_{i} ^{(\pi)}| \otimes \langle \phi_{i}^{(\nu)} | H  R_a
  \\
  && 
     \sum_{b} W_b^{(Z)} R^{(Z)}_b| \phi_j^{(\pi)} \rangle
     \otimes \sum_{c} W_c^{(N)} R^{(N)}_c | \phi_j^{(\nu)} \rangle . 
     \nonumber \\
  N_{iK,jK'}
  &=& \sum_{a} W_a^{JKK'\pi}
      \langle \phi_{i} ^{(\pi)}| \otimes \langle \phi_{i}^{(\nu)} | R_a
      \\
  && 
     \sum_{b} W_b^{(Z)} R^{(Z)}_b| \phi_j^{(\pi)} \rangle
     \otimes \sum_{c} W_c^{(N)} R^{(N)}_c | \phi_j^{(\nu)} \rangle . 
     \nonumber 
\end{eqnarray}
% obtained as 
% \begin{eqnarray}
%   E_{N_b} &=& \langle \Psi_{N_b} | H | \Psi_{N_b} \rangle
%   \\
%   &=& \sum_{nn'}^{N_b} \sum_{K,K'=-J}^{J}
%   (f^{(N_b)}_{n'K'})^* f^{(N_b)}_{nK}
%   \sum_{a} W_a^{JK'K\pi}
%   \nonumber \\
%   && 
%   \langle \phi_{n'} ^{(\pi)}| \otimes \langle \phi_{n'}^{(\nu)} | H  R_a
%   \nonumber \\
%   && 
%   \sum_{b} W_b^{(Z)} R^{(Z)}_b| \phi_n^{(\pi)} \rangle
%   \otimes \sum_{c} W_c^{(N)} R^{(N)}_c | \phi_n^{(\nu)} \rangle . 
%   \nonumber
% \end{eqnarray}
Note that $f_{iK}^{(N_b)}$ is determined so that
the resultant wave function is normalized.
The number of the mesh points of the angular-momentum, parity
projector typically reaches 60,000 and 
the matrix element for each $a$ can be computed in parallel.
This feature is suitable for massively parallel computations. 
Since the computational cost for such variation after projection
is quite heavy, we utilized state-of-the-art supercomputers in Japan 
such as Fugaku and Oakforest-PACS. 
The developed code is equipped with the code-tuning technique
suggested in Ref.~\cite{mcsm_tuning}.

Hereafter, we consider only protons for simplicity.
The derivation of its extension to
the proton-neutron system is lengthy but straightforward.
Since the rotated quasi-particle vacuum $R|\phi\rangle$
can be expressed as 
another quasi-particle vacuum
by using the Baker-Campbell-Hausdorff formula, 
we need to compute the matrix element of the Hamiltonian
between two different quasi-particle vacua. 
Note that the $f^{(N_b)}_{iK}$ is obtained by 
solving Eq.(\ref{eq:geneig}) every time its basis state is changed.

The Hamiltonian matrix element between the different
quasi-particle vacua, $|\phi\rangle$ and $|\phi'\rangle$, 
is calculated using the generalized Wick theorem
\cite{ringschuck}  as
\begin{eqnarray}
  \langle \phi| H |\phi' \rangle
  &=& \langle \phi | \phi' \rangle
  \textrm{Tr} \left(
  t \rho + \frac12\Gamma \rho - \frac12 \kappa'\Delta 
  \right)
\end{eqnarray}
where the density matrix $\rho$ and the pairing tensor $\kappa$
are obtained as
\begin{eqnarray}
  \rho_{ij}
  &=&
  \frac{\langle \phi|c^\dagger_j c_i |\phi' \rangle}{\langle \phi|\phi' \rangle }
  = -Z' (1-Z^* Z')^{-1}Z^*
  \\
  \kappa_{ij}
  &=&
  \frac{\langle \phi|c_j c_i |\phi'\rangle}{\langle \phi|\phi'\rangle }
  = Z' (1-Z^* Z')^{-1}
  \\
  \kappa'_{ij}
  &=&
  \frac{\langle \phi|c^\dagger_i c^\dagger_j |\phi'\rangle}{\langle \phi|\phi' \rangle }
  =  (1-Z^* Z')^{-1} Z^* 
\end{eqnarray}
\begin{eqnarray}
  Z &=&  (V U^{-1})^*
  \nonumber \\
  Z' &=& (V'U'^{-1})^*
\end{eqnarray}
%% \begin{eqnarray}
%%   |\phi \rangle &=& \exp\left(\sum_{i<j} Z_{ij}c^\dagger_i c^\dagger_j\right) | - \rangle
%%   \nonumber \\
%%   |\phi' \rangle &=& \exp\left(\sum_{i<j} Z'_{ij}c^\dagger_i c^\dagger_j\right) | - \rangle
%% \end{eqnarray}
\begin{eqnarray}
  \Gamma_{ik}
  &=&
  \sum_{jl} v_{ijkl} \rho_{lj}
  \nonumber \\
  \Delta_{ij} 
  &=&
  \frac12 \sum_{kl} v_{ijkl} \kappa_{kl} . 
\end{eqnarray}

\subsection{Energy gradient}

In the QVSM scheme, we apply the conjugate gradient method to minimize
the projected energy expectation value.
The basis state $|\phi_n\rangle$ is determined sequentially
so that the energy expectation value $E_{N_b=n}$ is minimized.
Let us consider the situation that $N_b-1$ basis states have already
been fixed and the variational calculation is performed 
with the variational parameters of the $N_b$-th basis state, $Z^{(N_b)}$, 
by the conjugate gradient method, which
requires the gradient of the projected energy
of the superposed quasi-particle vacua.
The energy gradient is obtained as
\begin{eqnarray}
  \frac{\partial E_{N_b}}{\partial Z^{(N_b)*}} 
  &=&
  \sum_{n,KK'} (f^{(N_b)}_{N_b K'})^* f^{(N_b)}_{n K}
  \frac{\partial \langle \phi_{N_b}| }{\partial Z^{(N_b)*}}
  \nonumber \\
  &&
  (H - E_{N_b})P^{J\pi}_{K'K}P^{Z} | \phi_n \rangle , 
\end{eqnarray}
where the matrix element of the gradient is obtained as
\begin{eqnarray}
  && 
  \frac{\partial \langle \phi| }{\partial Z^*} 
  (H -E_{N_b})| \phi' \rangle
%  \nonumber \\
%  &=&  \langle \phi|\beta_{j}\beta_{i} (H -E_{N_b})|\phi' \rangle
  \nonumber \\
  &=& \langle \phi|\phi'\rangle
  ( U_D^\dagger (t+\Gamma) V_D^*  - V_D^\dagger (t+\Gamma)^T U_D^*
  \nonumber \\
  && 
  + U_D^\dagger \Delta U_D^* - V_D^\dagger \Delta' V_D^*)
  \nonumber \\
  && 
  -  Z_D  \langle \phi'| (H-E_{N_b})| \phi \rangle 
\end{eqnarray}
with
\begin{eqnarray}
  Z_D &=& ( ( V^T U' + U^T V' )( U^\dagger U' + V^\dagger V' )^{-1})^*
  \nonumber \\
  U_D &=& U + V^* Z_D^*
  \nonumber \\
  V_D &=& V + U^* Z_D^* .
  \nonumber  \\
  \Delta'_{kl} &=&  \frac12 \sum_{ij} \kappa'_{ij}v_{ijkl}  .
\end{eqnarray}

\subsection{Energy variance}
\label{sec:app_variance}

Since the energy variance is the expectation value of the
four-body operator, 
the computation of the energy variance is 
time-consuming and its efficient computation is essential for practical
applications.
By utilizing the separability of $H^2$ in a similar way
to the case of Slater determinants in Ref.~\cite{mcsm_extrap},
a formula to compute the matrix element of the Hamiltonian
squared between two quasi-particle vacua is given as 
\begin{eqnarray}
  && \langle \phi| H^2 |\phi' \rangle
  \nonumber \\
  &=& \langle \phi | \phi' \rangle \Big(
  \frac14 \sum_{ijkl} (\rho' v \rho)_{ijkl} 
  (\rho v \rho')_{klij}
  \nonumber \\
  &&
  + \frac14 \sum_{ijkl} (\kappa'v \kappa)_{ijkl}
  ( \kappa' v \kappa)_{jilk}
  - \sum_{ijkl} (\rho' v \rho)_{ijkl}
  ( \kappa' v \kappa)_{jkli}
  \nonumber \\
  &&
  + \frac12 \textrm{Tr} \big(
  (  \rho\Gamma_t - \kappa \Delta') ( \rho'\Gamma_t  + \kappa \Delta')
  \nonumber \\
  && 
  \ \ \ \ \ \ + (\Gamma_t\rho - \Delta \kappa') ( \Gamma_t \rho' + \Delta \kappa')
  \nonumber \\
  && 
  \ \ \ \ \ \ - (\kappa \Gamma_t^T - \rho\Delta) ( \kappa'\Gamma_t - \rho^T \Delta') 
  \nonumber \\
  && 
  \ \ \ \ \ \ - (\kappa \Gamma_t^T + \rho'\Delta) ( \kappa'\Gamma_t  + \rho'^T\Delta')
  \big)
  \nonumber \\
  && 
  + \big( \textrm{Tr}( t \rho +  \frac12\Gamma \rho
  - \frac12 \kappa'\Delta ) \big)^2
  \Big)
  \label{eq:variance}
\end{eqnarray}
with
\begin{eqnarray}
  \rho'_{ij} &=& \delta_{ij} - \rho_{ij}
  \nonumber  \\
  (\Gamma_t)_{ij} &=&  \Gamma_{ij} + t_{ij}
  \nonumber  \\
  (\rho' v \rho)_{ijkl} 
  &=& \sum_{a,c} \rho'_{ia} v_{ajcl}  \rho_{ck}
  \nonumber  \\
  (\rho v \rho')_{ijkl}
  &=& \sum_{b,d} \rho_{jb} v_{ibkd}  \rho'_{dl} 
  \nonumber \\
  (\kappa' v \kappa)_{ijkl}
  &=& \sum_{ac} \kappa'_{ia} v_{ajcl} \kappa_{ck} .
  %% \nonumber \\
  %% (\kappa v \kappa')_{ijkl}
  %% &=& \sum_{b,d} \kappa_{jb} v_{ibkd}  \kappa'_{dl} .
\end{eqnarray}
The angular-momentum, parity, and number projections
can be applied in the same way as described
in Appendix \ref{sec:mele}.
The most time-consuming part in practical calculations is
to compute $(\rho' v \rho)_{ijkl}$ as 
\begin{equation}
  (\rho' v \rho)_{ijkl} = 
  \sum_{a} \rho'_{ia}
  \left( \sum_c v_{ajcl}  \rho_{ck} \right).
\end{equation}
This is computed by the summations of the fivefold loops,
which cost far smaller than the case of
a general four-body operator demanding eightfold loops.


\begin{thebibliography}{99}


\bibitem{caurier_rmp}
  E. Caurier, G. Martinez-Pinedo, F. Nowacki,
  A. Poves, and A. P. Zuker,
  Rev. Mod. Phys. \textbf{77}, 427 (2005).

\bibitem{otsuka_rmp}
  T. Otsuka, A. Gade, O. Sorlin, T. Suzuki, and Y. Utsuno
  Rev. Mod. Phys. \textbf{92}, 015002 (2020).

\bibitem{mfdn} J. P. Vary, P. Maris, E. Ng, C. Yang,
  and M. Sosonkina,
  J. Phys.: Conf. Ser. {\bf 180} 012083 (2009).

\bibitem{bigstick}
  C. W. Johnson, W. E. Ormand, and P. G. Krastev,
  Comp. Phys. Comm. \textbf{184}, 2761 (2013).

\bibitem{kshell}  N. Shimizu, T. Mizusaki, Y. Utsuno, and Y. Tsunoda, 
  Comp. Phys. Comm. \textbf{244}, 372 (2019).

\bibitem{proj_sm}
  K. Hara and Y. Sun, Int. J. Mod. Phys. E \textbf{4}, 637 (1995);
  Y. Sun, K. Hara, J. A. Sheikh, J G. Hirsch, V. Velazquez, and M. Guidry,
  Phys. Rev. C \textbf{61}, 064323 (2000).

\bibitem{npa}
  Y. M. Zhao and A. Arima, Phys. Rep. \textbf{545}, 1 (2014).

\bibitem{yoshinaga_npa}
  K. Higashiyama and N. Yoshinaga, 
  Phys. Rev. C \textbf{83}, 034321 (2011). 

\bibitem{mcsm_ppnp}  T. Otsuka, M. Honma, T. Mizusaki,
  N. Shimizu, and Y. Utsuno,
  Prog. Part. Nucl. Phys.  \textbf{47}, 319 (2001).

\bibitem{mcsm_ptep}
  N. Shimizu, T. Abe, M. Honma, T. Otsuka,
  T. Togashi, Y. Tsunoda, Y. Utsuno, and T. Yoshida,
  Phys. Scr. \textbf{92}, 063001 (2017);
  N. Shimizu, T. Abe, Y. Tsunoda, Y. Utsuno, T. Yoshida,
  T. Mizusaki, M. Honma, and T. Otsuka,
  Prog. Theor. Exp. Phys. \textbf{2012}, 01A205 (2012)

\bibitem{vampir}
  K. W. Schmid, F. Grummer, and A. Faessler,
  Annal. Phys. \textbf{180}, 1 (1987).

\bibitem{vampir_ppnp}
  K. W. Schmid, Prog. Part. Nucl. Phys.
  \textbf{46}, 145 (2001).

\bibitem{puddu_hmd} G. Puddu,
  J. Phys. G: Nucl. Phys. \textbf{46}, 115103 (2019).

\bibitem{iter_diag}
  D. Bianco, N. Lo Iudice, F. Andreozzi, A. Porrino, and F. Knapp, 
  Phys. Rev. C \textbf{88}, 024303 (2013).

\bibitem{corr_sm}
  L. F. Jiao, Z. H. Sun, Z. X. Xu, F. R. Xu
  and C. Qi, Phys. Rev. C \textbf{90}, 024306 (2014).

\bibitem{dmrg_sm} O. Legeza, L. Veis, A. Poves, and J. Dukelsky
  Phys. Rev. C \textbf{92}, 051303(R) (2015).
  
\bibitem{itsm_extrap}
  C. Stumpf, J. Braun, and R. Roth,
  Phys. Rev. C \textbf{93}, 021301(R) (2016).

%\bibitem{dstep} L. Coraggio, A. Gargano, and N. Itaco
  %  Phys. Rev. C  \textbf{93}, 064328 (2016)

\bibitem{hfb_gcm}
  B. Bally, A. S. Fernandez, T. R. Rodriguez,
  Phys. Rev. C \textbf{100}, 044308 (2019).

\bibitem{yao_gcm_srg_1}
  J. M. Yao, J. Engel, L. J. Wang, C. F. Jiao, and H. Hergert,
  Phys. Rev. C  \textbf{98}, 054311 (2018).

\bibitem{yao_gcm_srg_2}
  J. M. Yao, A. Bally, J. Engel, R. Wirth, T. R. Rodriguez, and H. Hergert, 
  Phys. Rev. Lett. \textbf{124}, 232501 (2020).

\bibitem{yao_gcm_srg_3}
  J. M. Yao, A. Bally, R. Wirth, T. Miyagi, C. G. Payne, S. R. Stroberg,
  H. Hergert, and J. D. Holt, arXiv:2010.08609

\bibitem{smmc}
  S. E. Koonin, D. J. Dean, and K. Langanke,
  Phys. Rep. \textbf{278}, 1 (1997).

\bibitem{smmc_gammasoft}
  Y. Alhassid, G. F. Bertsch, D. J. Dean, and S. E. Koonin,
  Phys. Rev. Lett. \textbf{77}, 1444 (1996).

\bibitem{cp_smmc}
  J. Bonnard and O. Juillet,
  Phys. Rev. Lett. \textbf{111}, 012502 (2013).

\bibitem{mcsm_extrap}
  N. Shimizu, Y. Utsuno, T. Mizusaki, T. Otsuka, 
  T. Abe, and M. Honma, 
  Phys. Rev. C  \textbf{82}, 061305(R) (2010);
  N. Shimizu, Y. Utsuno, T. Mizusaki, 
  M. Honma, Y. Tsunoda, and T. Otsuka, 
  Phys. Rev. C \textbf{85}, 054301 (2012). 

\bibitem{qmcd_otsuka}
  T. Otsuka, M. Honma, and T. Mizusaki,
  Phys. Rev. Lett. \textbf{81}, 1588 (1998).

\bibitem{ytsunoda_ni}
  Y. Tsunoda, T. Otsuka, N. Shimizu, M. Honma, and Y. Utsuno
  Phys. Rev. C \textbf{89}, 031301(R) (2014).

\bibitem{utsuno_ioi}
  Y. Utsuno, T. Otsuka, T. Mizusaki and M. Honma,
  Phys. Rev. C \textbf{60}, 054315 (1999).

\bibitem{ntsunoda_prc}
  N. Tsunoda, T. Otsuka, N. Shimizu, M. H.-Jensen,
  K. Takayanagi, and T. Suzuki,
  Phys. Rev. C \textbf{95}, 021304(R) (2017).

\bibitem{ntsunoda_nature}
  N. Tsunoda, T. Otsuka, K. Takayanagi, N. Shimizu,
  T. Suzuki, Y. Utsuno, S. Yoshida and H. Ueno,
  Nature \textbf{587}, 66 (2020).

\bibitem{liu_mcsm}
  L. Liu, T. Otsuka, N. Shimizu, Y. Utsuno
  and R. Roth,
  Phys. Rev. C \textbf{86}, 014302 (2012).

\bibitem{tabe_mcsm}
  T. Abe, P. Maris, T. Otsuka, N. Shimizu,
  Y. Utsuno, and J. P. Vary,
  Phys. Rev. C \textbf{86}, 054301 (2012).

\bibitem{togashi_zr}
  T. Togashi, Y. Tsunoda, T. Otsuka, and N. Shimizu,
  Phys. Rev. Lett. \textbf{117}, 172502	 (2016).

\bibitem{togashi_sn}
  T. Togashi, Y. Tsunoda, T. Otsuka, and N. Shimizu,
  Phys. Rev. Lett. \textbf{117}, 172502 (2016).
  
\bibitem{otsuka_sm}
  T. Otsuka, Y. Tsunoda, T. Abe, N. Shimizu, and P. Van Duppen,
  Phys. Rev. Lett. \textbf{123}, 222502 (2019).

\bibitem{bmpines}
  A. Bohr, B. R. Mottelson, and D. Pines,
  Phys. Rev. \textbf{110}, 936 (1958).

\bibitem{ba_mcsm}
  N. Shimizu, T. Otsuka, T. Mizusaki, and M. Honma, 
  Phys. Rev. Lett.  \textbf{86}, 1171 (2001).

\bibitem{egidoring}
  J. L. Egido and P. Ring, Nucl. Phys. {\bf A383} 189 (1982).

\bibitem{avignone08}
  F. T. Avignone III, S. R. Eliott, J. Engel,
  Rev. Mod. Phys. \textbf{80}, 481 (2008).

\bibitem{engel_menendez_bb_review}
  J. Engel and J. Menendez,
  Rep. Prog. Phys. \textbf{80}, 046301 (2017).

\bibitem{caurier_bb}
  E. Caurier, F. Nowacki, A. Poves, and J. Retamosa,
  Phys. Rev. Lett. \textbf{77}, 1954 (1996).
  
\bibitem{iwata_bb}
  Y. Iwata, N. Shimizu, T. Otsuka, Y. Utsuno,
  J. Menendez, M. Honma and T. Abe,
  Phys. Rev. Lett. \textbf{116}, 112502 (2016).

\bibitem{dgt_bb}
  N. Shimizu, J. Menendez, and K. Yako,
  Phys. Rev. Lett. \textbf{120}, 142502 (2018).

\bibitem{menendez2009}
  J. Menendez, A. Poves, E.Caurier, F.Nowacki,
  Nucl. Phys. A \textbf{818}, 139 (2009).

\bibitem{corragio_bb}
  L. Coraggio, A. Gargano, N. Itaco, R. Mancino, and F. Nowacki,
  Phys. Rev. C \textbf{101}, 044315 (2020).

\bibitem{ringschuck}
  P. Ring and P. Schuck, ``The Nuclear Many-Body Problem'',
  Springer (1980).

\bibitem{num_recipe}
  Numerical Recipes in Fortran 77,
  the Art of Scientific Computing, 1992
  (Cambridge University Press, Cambridge, UK), 2nd ed.

\bibitem{shapecoex_vampir}
  A. Petrovici, K. W. Schmid, A. Fassler,
  J. M. Hamilton and A. V. Ramayya,
  Prog. Part. Nucl. Phys. \textbf{43}, 485 (1999).
  

\bibitem{sn100pn}
   B. A. Brown, N. J. Stone, J. R. Stone, I. S. Towner, and M. Hjorth-Jensen, 
   Phys. Rev. C \textbf{71}, 044317 (2005).

\bibitem{sorella}
  S. Sorella, Phys. Rev. B \textbf{64}, 024512 (2001).

\bibitem{mizusaki_imada_extrap}
  T. Mizusaki and M. Imada, Phys. Rev. C \textbf{65}, 064319
  (2002); {\it ibid.} \textbf{67}, 041301(R) (2003).

\bibitem{puddu_variance}
  G. Puddu, J. Phys. G: Nucl. Part. Phys.  \textbf{39},
  085108 (2012).

\bibitem{extrap_ncsm}
  H. Zhan, A. Nogga, B. R. Barrett, J. P. Vary, and P. Navratil,
  Phys. Rev. C \textbf{69}, 034302 (2004).

\bibitem{vmcsm}
  N. Shimizu, and T. Mizusaki,
  Phys. Rev. C \textbf{98}, 054309 (2018),
  T. Mizusaki and N. Shimizu, 
  Phys. Rev. C \textbf{98}, 054309 (2018).  

\bibitem{haxton}
  W. C. Haxton and G.J. Stephenson Jr.,
  Prog. Part. Nucl. Phys. \textbf{12}, 409 (1984).

 \bibitem{jun45} M. Honma, T. Otsuka, T. Mizusaki, and M. Hjorth-Jensen,
   Phys. Rev. {\bf C 80}, 064323 (2009).

\bibitem{caurier2008}
  E. Caurier, J. Menendez, F. Nowacki, and P. Poves,
  Phys. Rev. Lett. \textbf{100}, 052503 (2008)

\bibitem{rodriguez_150Nd}
  T. R. Rodriguez and G. M.-Pinedo
  Phys. Rev. Lett. \textbf{105}, 252503 (2010).

 \bibitem{gcm_bb}
   C. F. Jiao, J. Engel, and J. D. Holt,
   Phys. Rev. C \textbf{96}, 054310 (2017).

 \bibitem{khhe}
   T. T. S. Kuo and G. Herling, US Naval Research Laboratory
   Report No. 2258, 1971 (unpublished), Nucl. Phys. A
   \textbf{181}, 113 (1972).
   
 \bibitem{Yao_PRC91}
   J. M. Yao, L. S. Song, K. Hagino, P. Ring, and J. Meng,
   Phys. Rev. C \textbf{91}, 024316 (2015).

 \bibitem{QRPA_CH}
   M. T. Mustonen and J. Angel,
   Phys. Rev. C \textbf{87}, 064302 (2013).

 \bibitem{YaoEngel_PRC94}
   J. M. Yao and J. Engel
   Phys. Rev. C \textbf{94}, 014306 (2016).

 \bibitem{song_RMF_150Nd}
   L. S. Song, J. M. Yao, P. Ring, and J. Meng
   Phys. Rev. C \textbf{90}, 054309 (2014).

 \bibitem{FangFaessler2010}
   D.-L. Fang, A. Faessler, V. Rodin,
   and F. Simkovic,
   Phys. Rev. C 82, 051301(R) (2010).

 \bibitem{kamland}
   A. Gando {\it et al.},
   Phys. Rev. C 86, 021601(R) (2012).

 \bibitem{cupid}
   E. Armengaud  {\it et al.},
   Euro. Phys. J. C \textbf{80}, 44 (2020).
   
 \bibitem{neergardwust}
   K. Neergard and E. Wust, Nucl. Phys. {\bf A402} 311 (1983).

 \bibitem{robledo_pfaffian}
   L. M. Robledo, Phys. Rev. C \textbf{79}, 021302(R) (2009).

 \bibitem{AvezBender}
   B. Avez and M. Bender
   Phys. Rev. C \textbf{85}, 034325 (2012).

 \bibitem{bertsch_robledo}
   G. F. Bertsch and L. M. Robledo
   Phys. Rev. Lett. \textbf{108}, 042505 (2012).
   
 \bibitem{oimizusaki_odd}
   M. Oi and T. Mizusaki,
   Phys. Lett. B \textbf{707}, 305 (2012).

 \bibitem{hillwheeler}
   J. J. Griffin and J. A. Wheeler, Phys. Rev. \textbf{80}, 367 (1966).
   
 \bibitem{mcsm_tuning}
   Y. Utsuno, N. Shimizu, T. Otsuka
   and T. Abe, Comput. Phys. Comm. \textbf{184}, 102 (2013).

   

\end{thebibliography}
\end{document}